\documentclass[12pt]{iopart}
\usepackage{graphicx}
\def\la{\mathrel{\mathchoice {\vcenter{\offinterlineskip\halign{\hfil
$\displaystyle##$\hfil\cr<\cr\sim\cr}}}
{\vcenter{\offinterlineskip\halign{\hfil$\textstyle##$\hfil\cr
<\cr\sim\cr}}}
{\vcenter{\offinterlineskip\halign{\hfil$\scriptstyle##$\hfil\cr
<\cr\sim\cr}}}
{\vcenter{\offinterlineskip\halign{\hfil$\scriptscriptstyle##$\hfil\cr
<\cr\sim\cr}}}}}
\def\ga{\mathrel{\mathchoice {\vcenter{\offinterlineskip\halign{\hfil
$\displaystyle##$\hfil\cr>\cr\sim\cr}}}
{\vcenter{\offinterlineskip\halign{\hfil$\textstyle##$\hfil\cr
>\cr\sim\cr}}}
{\vcenter{\offinterlineskip\halign{\hfil$\scriptstyle##$\hfil\cr
>\cr\sim\cr}}}
{\vcenter{\offinterlineskip\halign{\hfil$\scriptscriptstyle##$\hfil\cr
>\cr\sim\cr}}}}}
\def\url#1{{\ttfamily\def\/{/\discretionary{}{}{}}#1}}

\begin{document}

\title[Reionization history from CMB polarization]{Constraining the 
reionization history with large angle CMB polarization}
\author{Loris P L Colombo}
\address{Dipartimento di Fisica ``G. Occhialini'', Universit\`a di
  Milano-Bicocca, Piazza della Scienza 3, I20126 Milano, Italy}
\address{I.N.F.N., Via Celoria 16, I20133 Milano, Italy}
\ead{loris.colombo@mib.infn.it}
\begin{abstract}
The first-year WMAP data release showed that the reionization optical
depth to CMB photons is greater than previously thought. This 
follows from unexpectedly high values of the $C^{TE}_l$ spectrum, for
$l$ up to $\sim 30$, presumably allowing a measurement of the E-mode 
polarization spectrum $C^E_l$, sooner than expected.
This note aims to test the capability of large-angle polarization 
experiments to explore the history of the cosmological reionization, 
considering also the impact of cosmic variance. In particular, here we 
discuss how well the ionized fraction $x_e$ and the reionization
redshift $z_r$ can be separately measured, at various levels of
instrumental noise.
\end{abstract}


\submitto{JCAP}
\maketitle

\section{Introduction}
\label{sec:int}
In the last ten years, measures on Cosmic Microwave Background (CMB) 
have reached high sensitivity and resolution
(see, e.g., \cite{cobe,boom,max,dasi,map}),
providing the spectrum of temperature anisotropies, $C_l^T$, up to $l 
\sim 1000$. This allowed to better constrain various parameters of 
the cosmological model. Among them, the optical depth $\tau$, due 
to reionization, has a peculiar character. Other parameters directly
reflect initial conditions and/or inflationary outputs; the value
of $\tau$, instead, conveys information on the physical conditions in
the Universe due to more recent events, and opens an observational
window on the physics of ancient cosmic objects. Furthermore, while 
other parameters influence data through linear physics, the cosmic 
opacity $\tau$ is clearly linked to later non--linear evolution.

Unfortunately, $\tau$ is only mildly constrained by $C_l^T$, because 
of a degeneracy with the primordial spectral index ($n$) of density
fluctuations (\cite{kam,eft}; see also \cite{cb}). 
For this reason, pre--WMAP $C_l^T$ data could only constrain 
$\tau$ to be $\la 0.40$, at the 2--$\sigma$ level \cite{stom}.
On the other hand, several models of reionization hinted at a value
of the optical depth in the range $ 0.03-0.05$ 
(see, e.g, \cite{mir}).

The recent WMAP first--year release\footnote[1]{
\url{http://lambda.gsfc.nasa.gov/product/map/m\_overview.html}}
has significantly modified the
context, thanks to data on the $ET$--correlation spectrum,
$C_l^{ET}$. Quite in general, $C_l^{ET}$ and the $E$--polarization 
spectrum $C_l^E$ contain independent information on $\tau$. In
particular, the most part of this information is conveyed by the spectral
components with $l \la 30$; it is their unexpectedly
high amplitude that led to estimate $\tau 
\simeq 0.17\pm 0.04$ \cite{kogut}.

In the presence of such signal, it is legitimate to wonder whether
current (or more advanced) experiments can succeed in providing us
with more information on reionization, besides the value of $\tau$.
In this work we shall refer to the expected performance of the Sky 
Polarization Observatory
(SPOrt\footnote[2]{\url{http://sport.bo.iasf.cnr.it/}}
 \cite{sport,carr,macc}), 
first to outline its capacity to provide an independent confirm
of the high $\tau$ value, by using high--sensitivity polarimeters
operating at three MW frequencies (22, 32 and 90 GHz). Then, we
shall focus on the capability of low--$l$ polarization experiments
to inspect the cosmic reionization history, by analyzing the expected 
outputs of experiments run at increasing levels of sensitivity.
Our conclusions directly apply to any (nearly--)full sky experiment
with a design similar to SPOrt; using them for possible 
future experiment, with minor technical differences, is substantially
harmless.

At large angular scales, cosmic variance (CV) can never be neglected 
and risks to reduce the strength of CMB data in constraining
the reionization history. One of the main difficulties of this
analysis amounts to taking carefully into account its effects.
We shall see that, as these effects interfere with low--$l$ inspections, 
a sensitivity much higher than the one of WMAP or SPOrt ought to 
be reached, before the full potentiality of large angles results is
exploited.

\section{Monte Carlo analysis}

\subsection{Model properties}
\label{ssec:rem}
An optical depth $\tau =0.17 \pm 0.04$ corresponds to a reionization 
redshift $z_{r} \simeq 17 \pm 5$, assuming a sharp reionization in 
a $\Lambda$CDM model with WMAP best fit parameters. 
On the other hand, the Gunn--Peterson effect
seems to affect high-$z$ quasar radiation (\cite{djor,beck}), 
emitted at redshift $\simeq 6$; this shows that at least a fraction 
(possibly as small as $1\, \%$) of neutral hydrogen was still present 
at that time. Taken together, these observations imply that reionization
is a fairly complex process, and that its complete characterization requires
more than the value of the optical depth (\cite{ferr}).

To account for such an observational picture, several models have been
proposed, including double reionization models (\cite{cen,WL}) 
or a period of extended reionization
(\cite{extre}), but more complex pictures are possible. A feature
common to most these models is a bump in $C_l^E$ and $C_l^{TE}$  
spectra at low $l$, but other effects can be expected, involving 
spectral components up to $l >1000$.  
The analysis carried on in this work is based on a  partial
reionization model, in 
which baryonic materials reionize abruptly at a redshift $z_r$, 
attaining a reionization fraction $x_e \le 1$. Thus, a given value of 
$\tau$ corresponds to different $x_e-z_r$ pairs. Although this
is admittedly a toy model, it may reasonably approximate
a large family of possible reionization histories. We shall take
other relevant cosmological parameters as known from independent data
(e.g., from high--$l$ spectral analysis); we fix their values
as follows: a flat
spatial section with reduced density parameters $\Omega_m h^2 = 0.147$, 
$\Omega_b h^2 = 0.024 $ (for matter and baryons, respectively), the
Hubble parameter is 70 km/s/Mpc and the primeval spectral index is 
$n=1$. We assume no contributions from tensor modes of gravity
perturbation; this means, in particular, no $B$-mode polarization. 
In a previous work (\cite{cb}) we tested that, in practice, 
likelihood results are insensitive to reasonable variations of these
parameters.

Parameter estimation from CMB data usually follows a Bayesian maximum
likelihood approach (see, e.g., \cite{zald98,verde}). In this work,
however, we are mainly interested in polarization
measures at large angles, to test how they can improve the 
reconstruction of the reionization history. 
Therefore, we adopted a Monte Carlo approach to directly account for
CV, and performed a wide set of realizations of four 
models, corresponding to $\tau = 0.17$ or 0.13 with $x_e$ = 0.6 or 1.  

Artificial anisotropy and polarization data were built using
CMBFAST\footnote[7]{\url{http://physics.nyu.edu/matiasz/CMBFAST/cmbfast.html}}
and HEALPix\footnote[6]{\url{http://www.eso.org/science/healpix/}} codes.
The resulting maps have been smoothed with a purely Gaussian filter 
of $7^o$ FWHM, to represent SPOrt instruments beam, and we chose a HEALPix 
resolution 
$N_{side} = 16$, corresponding to a pixels' width of $\simeq 3.5^o$, thus
approaching the Nyquist frequency of the configuration.  
At the angular scales considered here, the non-ideality of SPOrt's feed horns 
give rise to a spurious polarization signal due mainly to the temperature 
anisotropy signal smoothed on scales equal to the instruments' FWHM.
Assuming a rms $T$ intensity of $\sim 30 \mu$K, this translates
into a rms contamination $< 0.2 \mu$K, which is roughly an order of
magnitude smaller than the expected noise of the SPOrt experiment 
\cite{sport,ettore}.    

We considered increasing levels of sensitivity for polarization 
(noise-per-pixel $\sigma_{pix} = 2$, 1, 0.5, 0.1$\, \mu$K). The first
value should be achieved by combining the two 90 GHz SPOrt-channels,
after two years of data taking, while using the 22 and 32 GHz channels
to map and remove the galactic foregrounds contamination.
Temperature noise was held fixed to $\sigma^T_{pix} =1 \mu$K, 
roughly the value expected for the 4-year WMAP results. 
As on large scales signal-to-noise ratio in temperature signal is much 
more than unity, varying temperature noise in the same range 
of $\sigma_{pix}$ does not affect our results. Noise was assumed to be
uncorrelated, both between different pixels and between the different
Stokes parameters. For each choiche of $\tau$, $x_e$ and $\sigma_{pix}$, 1000
independent sets of artificial data were considered.

\subsection{Likelihood evaluation}
\label{ssec:det}
When multi-parameter analysis for small angle experiments is
required, likelihood computation rapidly becomes time intensive,
forcing the adoption of various approximations (see, e.g.,
\cite{verde,bond}) or even a fisher-matrix approach. 
In this work, however, we are interested mainly in how polarization 
measures at large angles can improve the reconstruction of the 
reionization history. Moreover, SPOrt large beam size implies that 
data will be binned to a fairly low number of pixels. Together, these 
considerations make the evaluation of the full likelihood function in 
coordinate, or pixel, space feasible  for a reduced number of
parameters. This has the added advantage of simplifying the joint 
likelihood between experiments with different sky coverages.

An artificial CMB sky map is a realization of a given cosmological
model $\cal M$ whose statistical properties (i.e. the sets of $C^Y_l$
coefficients, with $Y=T,E,TE$) are known. On each pixel, measures of 
temperature fluctuations ($T$) and of the $Q$ and $U$ Stokes parameters 
are ideally performed. 

Galactic foregrounds contamination is a serious issue in $T$ maps, we 
therefore exclude the Galactic plane (i.e. the region of sky with 
Galactic declination $|\delta_g| <20^o$) from anisotropy maps.  
In the case of polarization, Galactic contamination, while still an issue,
plays a less important role. The first year WMAP data showed that the 
most prominent foreground contribution is due to Galactic synchrotron,
whose polarization spectrum has a steep dependency on frequency. 
According to the synchrotron template developed by the SPOrt team\cite{gianni},
at 90-100 GHz CMB polarization should dominate over foregrounds 
contamination by a factor of 2-3 to almost 1 order of magnitude, 
depending on the value of the optical depth.
Accordingly, we exclude from polarization maps only the equatorial polar caps, 
with  declination $|\delta| > 51.6^o$, which SPOrt will not be able 
to inspect from its position on board the ISS. Notice that the polar caps
partly overlap with the Galactic plane.
The number of pixels available for anisotropy (polarization) measures 
is therefore $N_T =1984$ ($N_P =2404$). These data form a 
$(N_T+2N_P)$-components vector 
${\bf x} \equiv \{T(i=1,...,N_T), Q(i=1,....,N_P), U(i=1,....,N_P)\}$.

The statistical properties of the process underlining the
synthetic data vector can be expressed through the correlation matrix
$\langle {\bf x^T}_i{\bf x}_j\rangle \equiv
{\bf C}_{ij}={\bf S}_{ij}+{\bf N}_{ij}$, where the brackets mean
ensemble average. In particular, the correlation matrix has been
written as the sum of a signal term ${\bf S}_{ij}$, which depends on
cosmology and geometry, and a term which accounts for detector
noise. 
 
The expected anisotropy correlation between 
two pixels with angular separation $\vartheta_{ij}$ is given by
\begin{equation}
\langle T_i T_j \rangle = \sum_l {2l + 1 \over 4\pi} C^T_l P_l (\cos
\vartheta_{ij}) B^2_l ;
\end{equation}
where $P_l (\cos\vartheta_{ij})$ are Legendre polynomials and $B^2_l$
are reduction coefficients due to pixelization and beam smoothing.  
For other correlations, similar expressions hold  and can be evaluated
using the relations given in \cite{zald98}.  

We assume no correlation among noise in different modes and pixels, so $\bf
N$$_{ij}$ is diagonal and has distinct values $(\sigma^T_{pix})^2$ and 
$\sigma_{pix}^2$ in the former $N_T$ and in the latter $2N_P$ pixels, 
respectively. 

Assuming Gaussian statistics, the likelihood of a model $\cal M'$
given the synthetic data {\bf x} reads then:
\begin{equation}
{\cal L}({\cal M'}|{\bf x}) = {1 \over (2\pi)^{N_T+2N_P}} 
{1 \over \sqrt{\det{\bf C'}}}
\exp \left(- {1 \over 2}{\bf x}^{\rm T}{\bf C'}^{-1}{\bf x}\right) ~.
\label{eq:like}
\end{equation}
In a Bayesian analysis, the posterior probability density is given by
the product of the likelihood function and the prior probability density.
According to Bayes theorem, in the case of an uniform prior the most
likely model is the one for whom the likelihood function is maximum. 
Confidence regions in parameter space are then found by a likelihood
ratio criterion. These regions are also called {\it credibility
 regions} and generally represent only an approximation to the true
confidence intervals, particularly when CV is significant.

A frequentist analysis, instead, stems from the question: What is
the probability of obtaining a given output from an observation if the
underlying physical process has a certain set of parameters?
To answer this question, the usual procedure is to generate a large set of
synthetic observations, analyze them as real data, and then study the
distribution of the results. Confidence regions, which in this case
are {\it exclusion regions}, are determined as the portion of
parameter space which contain a set percentage of the total results.

\section{Results}
\label{sec:res}
\begin{figure}
\begin{center}
\includegraphics*[width=9cm]{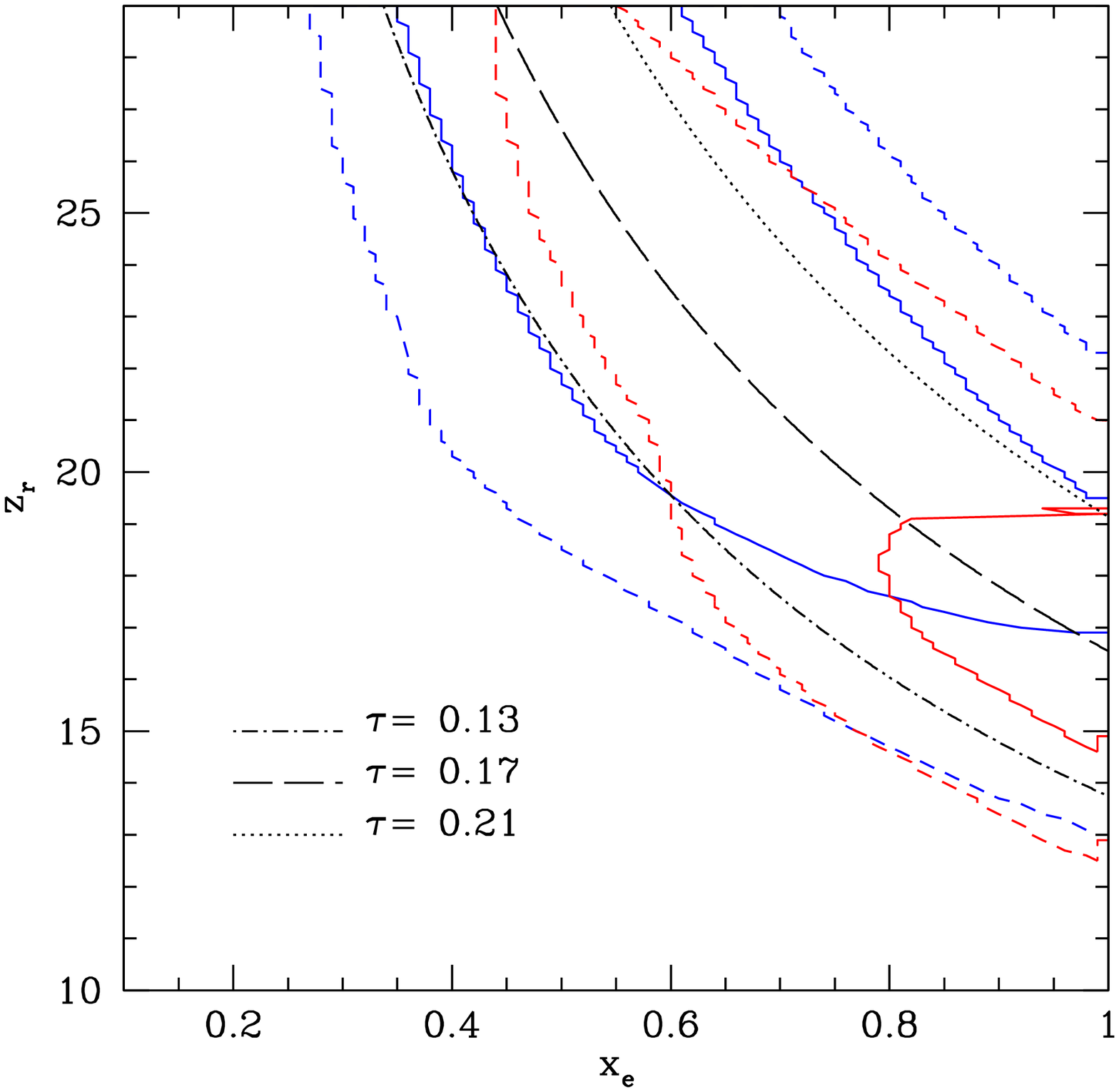}
\end{center}
\caption{\label{fig:s2} Expected likelihood contours in the $x_e-z_r$
  plane for models with $\tau =0.17$ and two different reionization
  histories; $\sigma_{pix} =2 \mu$K. Solid and short-dashed lines 
  trace the boundary of the 68\% and 95\% confidence level regions,
  respectively. In red we plot results for a model in which
  reionization is complete, while blue lines refer to a model in which
  $x_e = 0.6$. Likelihood contours are elongated along direction of
  constant optical depth end do not allow to distinguish between the
  two models.}
\end{figure}

\begin{figure}
\begin{center}
\vskip-2truecm
\includegraphics*[width=9cm]{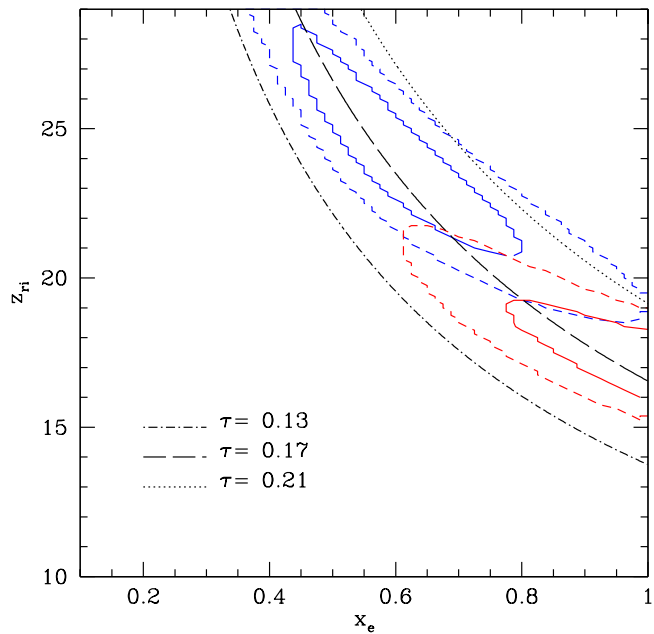}
\end{center}
\caption{\label{fig:s05} Same as \fref{fig:s2}, but with $\sigma_{pix}
  =0.5 \mu$K. Likelihood function are elongated along lines of
  constant $\tau$, it is possible to distinguish between the two
  models at least at 68\% c.l.}
\end{figure}

For each simulated sky, likelihood distributions have been evaluated 
on the $x_e$-$z_r$ plane. We show results in two ways: first, we
average over the whole set of 1000 realizations and show 1 and
2--$\sigma$ confidence regions (see \fref{fig:s2} and \fref{fig:s05}). 
Outside of them, we have {\it exclusion regions}, i.e. 
the portions of the $x_e$-$z_r$ that can be
excluded with a given degree of confidence, taking into account the
variance both in the detector noise and in the true CMB data.
For the range of $\sigma_{pix}$ we tested, with the exception of 
the lowest value, the detector noise variance 
however dominates polarization measures, while the proper CV is more
relevant for anisotropy data, where the signal to noise ratio is higher.

\begin{figure}
\begin{center}
\includegraphics*[width=9cm]{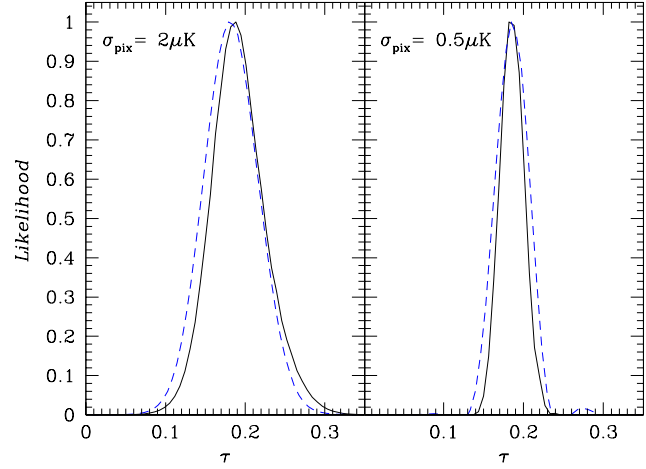}
\end{center}
\caption{\label{fig:tlike} Likelihood function after marginalizing
  over $x_e$, for the models of \fref{fig:s2}. Results are plotted 
  as a function of $\tau$ instead of $z_r$. Solid lines refer to
  complete reionization, dashed lines to $x_e =0.6$. No significant
  bias in the determination of the optical depth is introduced 
by the reionization histories considered here.}
\end{figure}

\begin{figure}
\begin{center}
\includegraphics*[width=9cm]{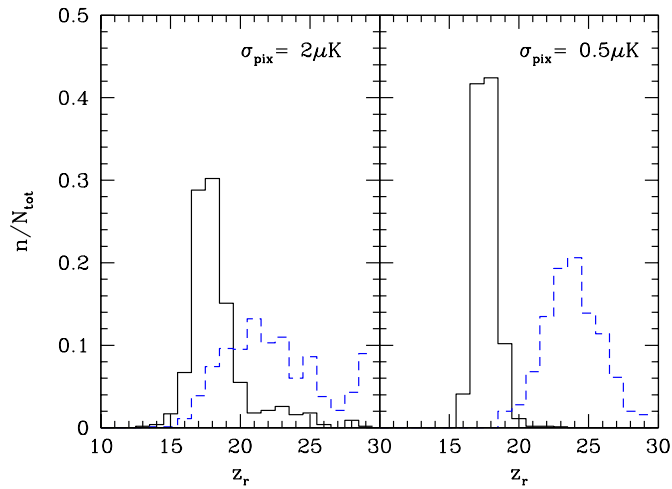}
\end{center}
\caption{\label{fig:zr} Histogram of the distribution of the maximum
  likelihood value of $z_r$, for models with $\tau =0.17$ and
  different reionization histories. Solid lines refer to $x_e =1, \, z_r
  \simeq 17$, dashed lines to $x_e =0.6, \, z_r \simeq 24$. Neat
  separation of the two distributions requires pixel noises $\sim 0.5 \mu$K.}
\end{figure}

\begin{figure}
\begin{center}
\includegraphics*[width=9cm]{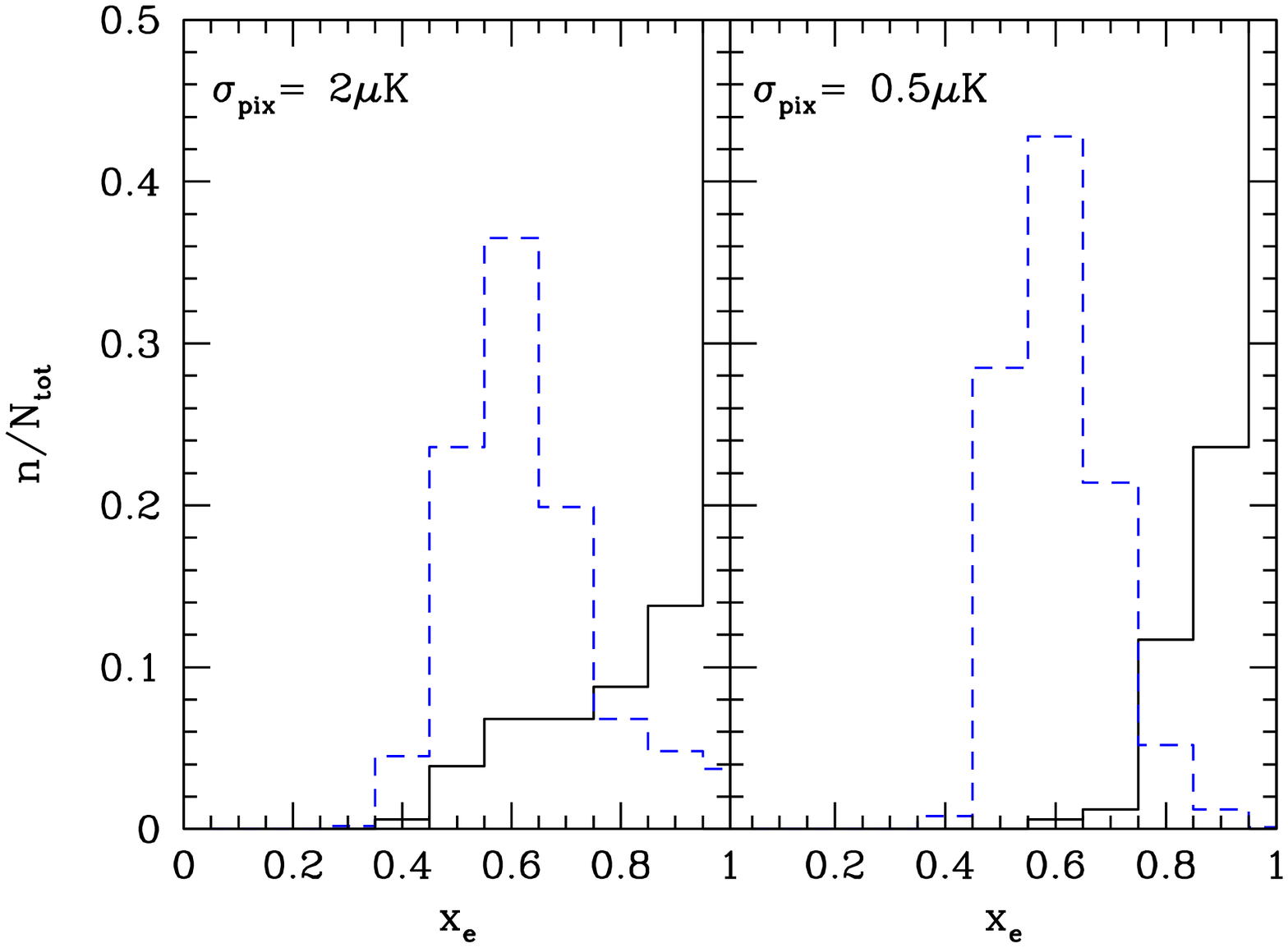}
\end{center}
\caption{\label{fig:xe} Same as \fref{fig:zr}, but for $x_e$.}
\end{figure}

However, when CV is significant, as it is for the angular separations 
considered here, the statistical properties of a realization can
differ significantly from those of the underlying process.
It is therefore significant to show how maximum likelihoods parameters 
are distributed. In \fref{fig:tlike}, the
distribution on $\tau$ is shown, both for $\sigma_{pix} = 2$
and 0.5$\, \mu$K; solid (dashed) lines refer to 
the case $x_e =1 ~~(= 0.6)$.  This plot is meant to test how far $\tau$
determinations are affected by the actual reionization history.
Then, in \fref{fig:zr}, we report the maximum likelihood
distribution over $x_e$, after marginalizing over $z_r$. Finally,
in \fref{fig:xe}, the distribution over $x_e$  is shown, after marginalizing 
over $z_r$. The same values of $\sigma_{pix}$ are considered, while
solid and dashed lines refer to the same cases as above.

\section{Discussion}
\label{sec:disc}
\Fref{fig:s2}  shows the 1 and 2--$\sigma$ confidence regions for
models with $\tau =0.17$, when $\sigma _P =2  \mu$K. 
They clearly show a likelihood distribution elongated along the
constant optical depth directions; this confirms that CMB data, 
for current noise levels, can provide constraints only on the
integrated value $\tau$ (see, e.g. \cite{brutau},\cite{kap}), but also
that such constraints are fairly safe. 

\Fref{fig:s05} shows the 1 and 2--$\sigma$ confidence regions, 
still for models with $\tau =0.17$, but when $\sigma_{pix} = 0.5  \mu$K. 
At this pixel noise level, the polarization sensitivity allows 
us to constrain an additional parameter ($x_e$ or $z_r$) beside $\tau$,
at least with a 68\% c.l. Once again, by marginalizing, e.g., over
$x_e$, we obtain the behaviour of the likelihood as a function of 
$\tau$. \Fref{fig:tlike} confirms that determination of the 
optical depth 
is substantially independent from the reionization history, at least
within the range of reionization models considered here (however,
other reionization histories can bias $\tau$ determinations 
\cite{hold}). This is
true both for $\sigma_{pix} = 2$ and 0.5$\, \mu$K.

By inspecting \fref{fig:zr}, we notice that, for the noise 
levels currently
achievable ($\sigma_{pix} \simeq 2\, \mu$K), the distributions of maximum 
likelihood values are greatly superimposed and it is not possible to 
distinguish between the two reionization histories. On the contrary,
for $\sigma_{pix} =0.5\, \mu$K, the superposition is restricted to 
the extreme tails of histograms; accordingly, at such a noise level,
some details on the reionization history can be inspected. Notice that
these plots provide the probability at which a given model can be {\it
  excluded} by the available data, instead of telling us the
likelihood of a given model being the correct one, as is instead done
in Bayesian analysis.

For both sensitivities, however, the distribution corresponding to
$x_e =1$ is more sharply defined than the one corresponding to
$x_e =0.6$. This is due to the fact that in the former case the true 
model necessarily lies on the boundary of the portion of parameter 
space considered. Analogous considerations could be made if,
instead of marginalizing on the reionization fraction, we had
marginalized over $z_r$, as it is shown in \fref{fig:xe}.  

As outlined above, in this work we inspected a number of further cases,
for which no plot is given. In particular, models with $\tau =0.13$ 
show quite a similar behavior, although some difference, on the level 
of noise required to achieve a clean separation of the two models,
exists. In particular, the overlap between maximum likelihood
histograms, for $\sigma_{pix} = 0.5 ~\mu$K worsens from 2$ \%$ to
$4.5\, \%$. For all cases, considering $\sigma_{pix} = 1\, \mu$K yields
intermediate results.

\begin{figure}
\begin{center}
\includegraphics*[width=9cm]{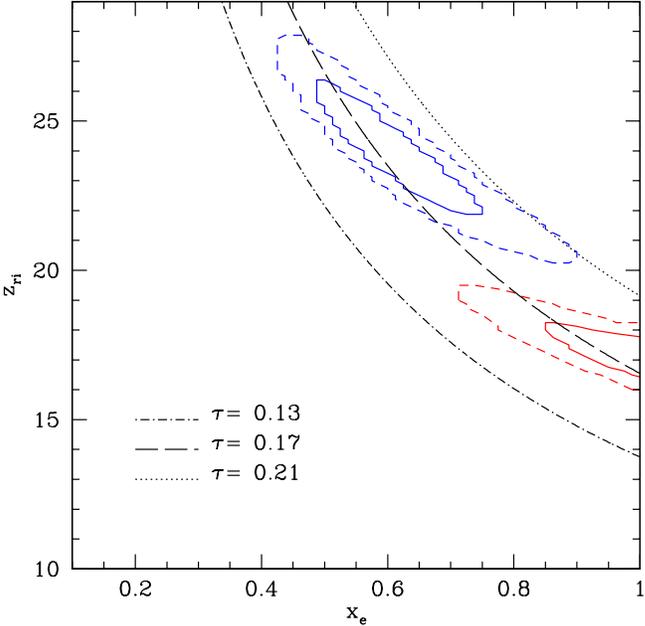}
\end{center}
\caption{\label{fig:s01} Same as \fref{fig:s2}, but with $\sigma_{pix}
  =0.1 \mu$K.  Clean separation between the two
  models is achieved at better than 95\% c.l.}
\end{figure}

We report here also results obtained for 
$\sigma_{pix} = 0.1\, \mu$K.
While this noise level is certainly beyond the capabilities of 
current instruments, it should be achieved by the 
PLANCK\footnote[3]{\url{http://astro.estec.esa.nl/Planck}}
experiment. While much of the information on reionization histories 
considered here is carried by the low-multipoles, PLANCK's higher 
spatial and frequency resolutions provide an advantage in the form of reduced 
sidelobes contamination and a better monitoring of foregrounds.  
Simulations show that clean separation between the two different 
reionization histories should be achieved at more 
than 95\% c.l., even taking into account CV (see \fref{fig:s01}). 
Moreover, at these noise levels, contribution from $B$-mode polarization 
should no longer be ignored, and could provide additional constraints 
on reionization. These, too, should be detected by the PLANCK satellite.

\section{Conclusions}
\label{sec:conc}
After the WMAP first-year results, it has become clear that
reionization history is a complex process and CMB measures can provide
unique information about it. In particular, CMB polarization (and, to
a lesser extent, temperature-polarization cross correlations) encode a
great deal of information not only on the total optical depth, but also on the
evolution with time of the ionized fraction. However, this knowledge
is found mainly in the first $l \la 30$ multipoles, i.e. at large
angles. These scales are strongly affected by cosmic variance;
therefore, the statistical properties of the actual sky can differ
significantly from those of the underlying cosmological model. 

In this work we investigated whether measures of CMB power spectra at
large angular separations can provide actual constraints on
reionization history. In particular, we considered the SPOrt
project as a benchmark of a large angles polarization measurements,
though we also explored instrumental sensitivities better than those
of the actual experiment. We considered toy models with the same
optical depth, but different fractions of ionized baryons; in both
case we assumed an instantaneous reionization. 

Using a frequentist analysis, we confirmed that, at the noise levels
currently achievable, CMB data can provide informations mainly on the
total optical depth. The SPOrt experiment will therefore provide the
first independent confirmation of WMAP results on the optical depth,
in a manner completely free of any spurious leakage between the $T$
and $E$ signals. 

Distinguishing between a model in which all baryons
reionized from a model in which the ionized fraction reaches only 60\%
requires, instead, noise levels significantly lower. In particular we
found that, for $\tau =0.17$, discriminating between the two models
at 68\% c.l. requires a polarization pixel noise $\sigma_{pix} = 0.5
 \mu$K, while for  $\ga 95\%$ c.l. $\sigma_{pix} \sim 0.1 \, \mu$K is needed. 
While the models of reionization considered here are admittedly
simplified, they provide two extreme examples of behaviour while being
consistent with the currently available data. Moreover, our main aim
was the determination of the sensitivity needed to extract
information on an additional reionization parameter from CMB data.
A complete assessment of the effects of a generic reionization history 
on CMB power spectra is certainly beyond the scope of this note, as 
it requires a completely different approach. In particular 
the non-uniform, or patchy, nature of the reionization process should 
be considered. This usually implies additional features at multipoles 
$l \ga 1000$, which need to be inspected in order to fully exploit 
the information on reionization encoded in CMB.      

Nonetheless, our results confirm that large angles polarization data 
can be used to discriminate between different reionization histories,
but that next generation CMB (polarization) experiments will be
indispensable for shedding light on those details of the reionization
process that can be inspected through this observational window.

\ack This work was financially supported by ASI, within the activities 
related to the SPOrt experiment. The author wishes to thank
S A Bonometto and the members of the SPOrt team for useful discussion 
and assistance in the preparation of the manuscript. Some of the results 
in this paper have been derived using the CMBFAST and HEALPix \cite{heal} packages.

\newpage

\end{document}